\documentstyle[12pt,epsf,epsfig]{article}
\begin{document}
\pagestyle{empty}
\begin{flushright}
CERN--TH/97--174\\
DAMTP--97--46
\end{flushright}

\vspace*{0.3cm}

\begin{center}
{\bf THE THREE-POINT FUNCTION IN SPLIT DIMENSIONAL REGULARIZATION IN THE
COULOMB GAUGE}\\
\vspace*{1cm}

{\bf G. Leibbrandt}\\
Department of Applied Mathematics and Theoretical Physics,\\
 University of Cambridge, Silver Street, Cambridge,  CB3 9EW, UK\\
and\\
Theoretical Physics Division, CERN, \\
CH-1211 Gen\`eve 23, Switzerland\\
\vspace*{1cm}

ABSTRACT
\end{center}
We use a gauge-invariant regularization procedure, called {\it split
dimensional regularization}, to evaluate the
quark self-energy $\rm \Sigma (p)$ and quark-quark-gluon vertex function
$\rm\Lambda_\mu (p^\prime,p)$ in
the Coulomb gauge, $\rm \vec{\bigtriangledown}\cdot\vec{A}^a = 0$. The
technique of split dimensional regularization was
designed to regulate Coulomb-gauge Feynman integrals in non-Abelian
theories. The technique which is based
on two complex regulating parameters, $\omega$ and $\sigma$, is shown to
generate a well-defined set of
Coulomb-gauge integrals. A major component of this project deals with
the evaluation of four-propagator and
five-propagator Coulomb integrals, some of which are {\it nonlocal}. It
is further argued that the standard
one-loop BRST identity relating $\Sigma$ and $\Lambda_\mu$, should by
rights be replaced by a more general
BRST identity which contains two additional contributions from ghost
vertex diagrams.
Despite the appearance of nonlocal Coulomb integrals, both $\Sigma$ and
$\Lambda_\mu$ are local functions
which satisfy the appropriate BRST identity.
Application of split dimensional regularization to two-loop energy 
integrals is briefly discussed.

\vspace*{1cm}
\noindent
Permanent address: On leave from the Department of Mathematics and
Statistics,
University of Guelph, Guelph, Ontario NIG 2W1.\\
E-mails: G.Leibbrandt@damtp.cam.ac.uk --
mthgeorg@mail.cern.ch
\vspace*{1cm}
\begin{flushleft}
CERN--TH/97--174\\
DAMTP--97--46\\
July 1997
\end{flushleft}
\vfill\eject

\pagestyle{plain}
\setcounter{page}{1}
\section{Introduction}

In our first test of the technique of {\it split dimensional
regularization}, we evaluated the Yang-Mills
self-energy to one loop and verified the appropriate BRST identity
\cite{1}. In this article, we present a
second test of split dimensional regularization, also in the Coulomb
gauge, by calculating the quark
self-energy and quark-quark-gluon vertex functions. Our rationale for
concentrating on one-loop
calculations at this stage is very simple: we want to make sure that
split dimensional regularization is
capable of regulating unambiguously all one-loop integrals in the
Coulomb gauge, before tackling the infamous
divergences at two loops and beyond.

The Coulomb gauge has been highly successful in both electrostatics and
quantum electrodynamics, especially in the
treatment of bound-state problems \cite{2}. By contrast, its application
to non-Abelian models has been marred by the
persistent appearance of ambiguous Coulomb-gauge Feynman integrals. It
was Schwinger \cite{3} who first noticed that
the transition from a classical Hamiltonian to a quantum Hamiltonian led
to instantaneous Coulomb-interaction terms.
These terms, subsequently labelled $\rm V_1+V_2$-terms by Christ and Lee
\cite{4}, are caused by the {\it operator
ordering problem}, and arise whenever we convert a classical system to a
quantum system. The principal problem with
these new
$\rm V_1+V_2$-terms is their nonlocality, foreshadowing potential
difficulties in evaluating Feynman integrals.

A review of the historical development of the Coulomb gauge in the
framework of Yang-Mills theory can be found in Ref.
\cite{2} and will not be repeated here. Suffice it to say, however, that
the nonlocal $\rm V_1+V_2$-terms of Christ and
Lee lead in general to momentum integrals containing two types of
divergences: (a) ordinary UV divergences, related to
the structure of space-time, and (b) new divergences, arising from the
integration over the energy variable $\rm dq_0$,
where $\rm d^4q \equiv dq_0~d^3 \vec{q}$. Hence the name {\it
energy-integrals} \cite{5,6}. It is the divergences from
the energy-integrals that cause the notorious ambiguities in the Coulomb
gauge, especially at two and three loops. To
compound the issue, standard dimensional regularization is not powerful
enough to consistently regulate the two types
of infinities mentioned above, as first pointed out by Doust and Taylor
\cite{7,8}.

In 1996, we introduced a radically different method, called {\it split
dimensional regularization}, for regulating
Feynman integrals in the Coulomb gauge and applied it to the one-loop
gluon self-energy $\rm \Pi^{ab}_{\mu\nu}$
\cite{1,9,10}. The new technique employs two complex-dimensional
parameters $\omega$ and $\sigma$, replacing the
measure $\rm d^4q = dq_0~d^3 \vec{q}$ by
\begin{equation}
\rm d^{2(\omega+\sigma)}q = d^{2\sigma} q_0~d^{2\omega} \vec{q}~,
\end{equation}
where the limits $\sigma \to (1/2)^+$, and $\omega \to (3/2)^+$ are to
be taken after all integrations have been
completed. Application of this procedure to the gluon self-energy
$\rm\Pi^{ab}_{\mu\nu} (p)$ leads to the following
results \cite{1}:
\begin{itemize}
\item[(i)] $\rm \Pi^{ab}_{\mu\nu}$ is local and nontransverse in the
Coulomb gauge.
\item[(ii)] Ghosts play a significant role, despite the physical nature
of the Coulomb gauge.
\item[(iii)] The BRST identity contains a nonvanishing ghost diagram,
but is rigorously satisfied.
\end{itemize}

Encouraging as this initial result may be, it would be foolhardy to
assume that split dimensional regularization can
regulate not only the remaining one-loop integrals in the Coulomb gauge,
but two- and three-loop integrals as well.
After all, it is well known how annoyingly unpredictable noncovariant
gauges may be at times! For this reason, we have
decided to proceed with the investigation of one-loop diagrams,
computing both the quark self-energy and
quark-quark-gluon vertex function.\\
\noindent
The latter is particularly challenging, since it contains for the first
time, Coulomb-gauge integrals with {\it five}
propagators.

The paper is organized as follows. In Section 2, we review the Feynman
rules in the Coulomb gauge, and evaluate the
quark self-energy $\Sigma$(p) using split dimensional regularization.
The quark-quark-gluon vertex function $\rm
\Lambda_\mu (p^\prime, p)$ is calculated in Section 3. In Section 4, we
illustrate the technique of split dimensional
regularization for more complicated Coulomb-gauge integrals: for a
massless four-propagator integral containing the
instantaneous Coulomb-interaction propagators $\rm (\vec{q}^2)^{-1}$ and
$\rm [(\vec{q}+\vec{q})^2]^{-1}$, and for a
massive five-propagator integral with propagators $\rm (\vec{q}^2)^{-1}$
and $\rm [(\vec{q}-\vec{k})^2]^{-1}$. In
Section 5, we examine the ghost contributions to the BRST identity and
verify the latter.  Implications of the technique of split dimensional
regularization for two-loop energy integrals are discussed in Section 6.
 The highlights of our
calculation are summarized in Section 7. The Appendix contains a partial
list of integrals needed for the determination
of $\Sigma$(p) and $\rm \Lambda_\mu (p^\prime,p)$.

\section{Feynman rules and computation of quark self-energy}

\subsection{Review of Feynman rules [11,12,13]}

In Yang-Mills theory, with the Lagrangian density
\begin{equation}
\rm L^\prime = L -\frac{1}{2\alpha} ({\it F}^{ab}_{\mu} A^{b\mu})^2~,
\alpha \equiv gauge~parameter,~\alpha \to 0~,
\end{equation}
where
\begin{eqnarray}
F{\rm ^{ab}_\mu} \!\! &\equiv&\!\!\rm (\partial_\mu -
\frac{n\cdot\partial}{n^2} n_\mu) \delta^{ab}~, ~~~~~\mu =
0,1,2,3~,
\nonumber\\
F{\rm ^{ab}_\mu A^{\mu b}}\!\!& =&\!\!\rm \vec{\bigtriangledown} \cdot
\vec{A}^a~,~~~~~n_\mu \equiv (n_0,\vec{n}) =
(1,\vec{0}), n^2 = 1~,\nonumber\\
\rm L \!\!&=&\!\! \rm -\frac{1}{4} (F^a_{\mu\nu})^2 + (J^c_\mu
+\bar{\omega}^a {\it F}^{ac}_\mu) D^{cb\mu} \omega^b
-\frac{1}{2} g f^{abc} K^a \omega^b \omega^c~,\nonumber\\
\rm F^a_{\mu\nu} \!\! &= & \!\!\rm \partial_\mu A^a_\nu -\partial_\nu
A^a_\mu + g f^{abc} A^b_\mu A^c_\nu~,\nonumber\\
\rm D^{ab}_\mu \!\! &=& \!\! \rm \delta^{ab} \partial_\mu + g f^{abc}
A^c_\mu~,\nonumber
\end{eqnarray}
the noncovariant Coulomb gauge is given by
\begin{equation}
\rm \vec{\bigtriangledown} \cdot \vec{A}^a (x) = 0~;
\end{equation}
$\rm A^a_\mu$ is a massless gauge field, g the gauge coupling constant,
$\rm f^{abc}$ are group structure constants,
and $\rm a = 1, ..., N^2-1$, for SU(N); $\rm \omega^a, \bar{\omega}^a$
represent ghost, anti-ghost fields
respectively, while $\rm K^a$ and $\rm J^a_\mu$ denote external sources.
The quantities $\rm J^a_\mu, \omega^a$ and
$\rm \bar{\omega}^a$ are anti-commuting.

In the Coulomb gauge, the Lagrangian density L$^\prime$ leads to the
following gauge boson propagator \cite{1}:
\begin{equation}
\rm G^{ab}_{\mu\nu} ({\it q}) =
\frac{-i\delta^{ab}}{(2\pi)^4(q^2+i\varepsilon)} \left[g_{\mu\nu} -
\left(\frac{n^2q_\mu
q_\nu -q
\cdot n (q_\mu n_\nu + q_\nu
n_\mu)}{-\vec{q}^2}\right)\right]~,~~~~~\varepsilon > 0~,
\end{equation} 
while the three-gluon vertex is given by
\begin{eqnarray}
\rm V^{abc}_{\mu\nu\rho} (p,{\it q},r) \!\! &= & \!\! \rm g f^{abc}
(2\pi)^4 \delta^4 (p+q+r) [g_{\mu\nu} (p-q)_\rho
\nonumber\\
 &+ & \!\! \rm g_{\nu\rho} (q-r)_\mu + g_{\rho\mu} (r-p)_\nu]~,
\end{eqnarray}
and the scalar ghost propagator by
\begin{equation}
\rm G^{ab}_{ghost} = \frac{i \delta^{ab}}{(2\pi)^4 \vec{q}^2}~.
\end{equation}

\subsection{Quark self-energy}

The unintegrated expression for the quark self-energy in Minkowski space
reads (cf. Fig. 1)
\begin{eqnarray}
\rm  \Sigma^{Coul} (p) \!\!&= &\!\! \rm \frac{4ig^2}{3}
\int\frac{d^4q}{(2\pi)^4 (q^2+i\varepsilon)} \gamma_\nu
\frac{1}{(\not{p}-\not{\it q}-m+i\varepsilon)} \gamma_\mu \nonumber\\ 
&& \!\!\rm  \left[ g^{\mu\nu}-\left( \frac{n^2q^\mu
q^\nu-q\cdot n(q^\mu n^\nu+q^\nu n^\mu)}{-\vec{\it
q}^2}\right)\right],~n_\mu = (1,0,0,0), \varepsilon > 0~,
\end{eqnarray}
where m is the quark mass; the generators $\rm T^a_{\alpha\beta}$ have
already been multiplied out. Expression (7)
leads to trivial covariant integrals, and to noncovariant Coulomb-gauge
integrals involving the factor $\rm
1/\vec{q}^2$. By performing a Wick rotation and applying split
dimensional regularization to all Coulomb integrals, we
may derive the following formulas in Euclidean space:

\setcounter{equation}{0}
\def\theequation{8\alph{equation}}
\begin{eqnarray}
&\rm div & \!\! \rm \int\frac{dq}{[(p-q)^2 + m^2]\vec{q}^2} =
2I^*~,~~~dq \equiv
\frac{d^{2(\sigma+\omega)}{\it q}}{(2\pi)^{2(\sigma+\omega)}}~, \\
&\rm  div & \!\!  \rm \int\frac{dq~{\it q}_i}{[(p-q)^2 +m^2]\vec{q}^2} =
\frac{2}{3} p_i I^*~,~~~i = 1,2,3,\\
&\rm  div & \!\! \rm \int \frac{dq}{q^2[(p-q)^2+m^2]\vec{q}^2} =
\frac{-2}{p^2+m^2} I^*~, \\
&\rm  div  & \!\! \rm \int \frac{dq~{\it q}_i{\it q}_j}{q^2[(p-q)^2+m^2]
\vec{q}^2} = \frac{1}{3} \delta_{ij} I^*~,
\end{eqnarray}

\def\theequation{\arabic{equation}}
\setcounter{equation}{8}
where
\begin{eqnarray}
\rm I^* & \equiv& \rm div \int \frac{d^{2\omega}\vec{\it
q}}{(2\pi)^{2\omega}} \int
\frac{d^{2\sigma}{\it q}_4}{(2\pi)^{2\sigma}}
\frac{1}{q^2 (q+p)^2}~,\nonumber\\
 & = & \rm divergent~part~of~\frac{\Gamma(2-\omega-\sigma)
(p^2)^{\omega+\sigma-2}}{(4\pi)^{\omega+\sigma}}~,\nonumber\\
& = & \rm \frac{\Gamma (2-\omega-\sigma)}{(4\pi)^2}~,~~\omega \to
(3/2)^+~, \sigma \to (1/2)^+~.
\end{eqnarray}

\begin{figure}[H]
%\centering
%\includegraphics[width=7cm]{(G.L)1.eps}
\hglue5cm
\epsfig{figure=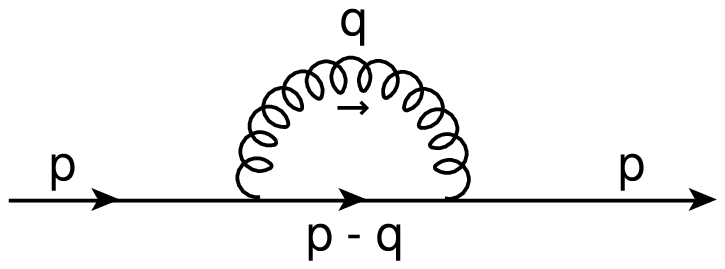,width=7cm}
\caption{Quark self-energy graph. Solid lines denote quarks, curly lines
gluons.}
\end{figure}
\noindent
Executing the required integrations in Eq. (7), we obtain
\begin{equation}
\rm \Sigma^{Coul}(p) = \frac{1}{3\pi} \alpha_s (\not{p}-4m)
\Gamma(2-\omega-\sigma)~,
\end{equation}
with $\rm \alpha_s \equiv \bar{g}^2/4\pi~, g \equiv
\bar{g}\mu^{2-\omega-\sigma}$, and $\mu$ being the mass scale. We
see that the result in Eq.~(10) is strictly {\it local}, despite the
appearance of nonlocal integrals such as Eq. (8c).

\section{Quark-quark-gluon vertex}

The quark-quark-gluon vertex function $\rm \Lambda_\mu (p^\prime,p)$
consists of the QED-like quark-quark-gluon vertex,
$\rm \Gamma^1_\mu (p^\prime, p)$, and the non-Abelian quark-quark-gluon
vertex, $\rm \Gamma^2_\mu (p^\prime,p)$:
\begin{equation}
\rm \Lambda_\mu (p^\prime,p) = \Gamma^1_\mu (p^\prime,p) + \Gamma^2_\mu
(p^\prime, p)~.
\end{equation}
$\Gamma^1_\mu$ (Fig. 2) is the easier one to compute, containing as it
does only the single noncovariant factor $\rm
1/\vec{q}^2$. Thus, 
\begin{eqnarray}
\rm \Gamma^1_\mu (p^\prime,p) \!\! & = & \rm  \!\!
\frac{ig^2}{6(2\pi)^4} \int
\frac{d^4q~\gamma_\rho~(\not{p}^\prime-\not{\it q}+m)\gamma_\mu
(\not{p}-\not{\it q}+m)
\gamma_\sigma}{q^2[(p^\prime-q)^2-m^2][(p-q)^2-m^2]} \nonumber\\
&&\nonumber\\
&& \!\! \rm \cdot
\left[g^{\rho\sigma}-\left(\frac{n^2{\it q}^\rho n^\sigma-q\cdot n({\it
q}^\sigma
n^\rho+{\it q}^\rho n^\sigma)}{-\vec{\it q}^2}\right)\right]~,
\end{eqnarray}
where the i$\varepsilon$-terms in the covariant propagators have been
omitted for clarity. The divergent component of
$\Gamma^1_\mu$ is, again, amazingly simple:
\begin{equation}
\rm \Gamma^1_\mu (p^\prime,p) = \frac{\alpha_s}{24\pi} \gamma_\mu \Gamma
(2-\omega-\sigma),~ \omega \to (3/2)^+, \sigma
\to (1/2)^+~.
\end{equation}
In deriving this answer, we have made use of the formulas listed in the
Appendix. Notice in particular the nonlocal
integrals there.

\begin{figure}[H]
%\centering
%\includegraphics[width=4cm]{(G.L)2.eps}
\hglue6cm
\epsfig{figure=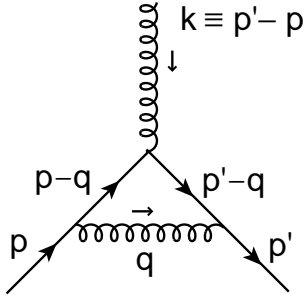,width=4cm}
\caption{QED-like quark-quark-gluon vertex graph.}
\end{figure}

Computation of the non-Abelian vertex function $\rm \Gamma^2_\mu
(p^\prime,p)$ in Fig. 3,
\begin{eqnarray}
\rm \Gamma^2_\mu (p^\prime,p) \!\!& =& \!\! \rm \frac{i3g^2}{2(2\pi)^4}
\int \frac{d^4q \gamma_\sigma
(\not{p}+\not{\it q}+m)\gamma_\nu}{q^2(k-q)^2[(p+q)^2-m^2]}
  \cdot \{g^{\rho\sigma} + \frac{1}{\vec{q}^2} [n^2  q^\rho q^\nu
-q\cdot n (q^\rho n^\nu +  q^\nu
n^\rho)]\}\nonumber\\
&& \!\! \rm \cdot \left\{ g^{\lambda\sigma} + \frac{1}{(\vec{k}-\vec{\it
q})^2} [n^2 (k-q)^\lambda (k-q)^\sigma - n\cdot
(k-{\it q})[(k-{\it q})^\lambda n^\sigma + (k-q)^\sigma
n^\lambda]]\right\} \nonumber\\
&& \!\! \rm \cdot [-(k+q)_\lambda g_{\mu\rho} + (2q-k)_\mu
g_{\rho\lambda} + (2k-{\it q})_\rho g_{\mu\lambda}]~, k
\equiv p^\prime-p~,
\end{eqnarray}
is complicated by the presence of the two noncovariant factors $\rm
1/\vec{q}^2$ and $\rm 1/(\vec{k}-\vec{q})^2$, and
the necessity of having to evaluate for the first time {\it
five}-propagator integrals (see Section 4 for details). In
fact, determination of the pole parts of these five-propagator integrals
in the Coulomb gauge took as long as the rest
of the calculation.

\begin{figure}[H]
%\centering
%\includegraphics[width=5cm]{(G.L)3.eps}
\hglue6cm
\epsfig{figure=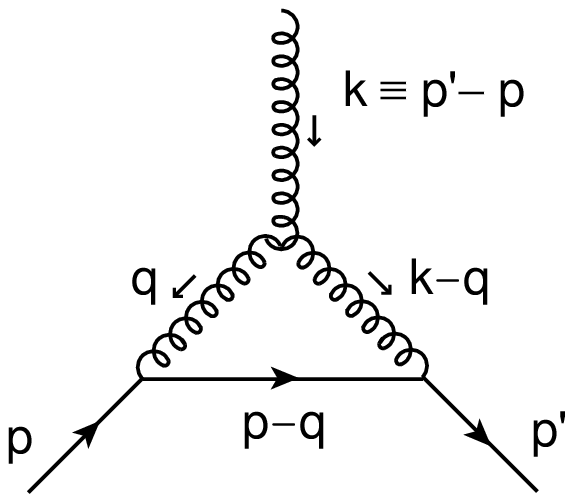,width=5cm}
\caption{Non-Abelian quark-quark-gluon vertex graph.}
\end{figure}

Despite the presence of explicitly nonlocal integrals in the evaluation
of $\rm \Gamma^2_\mu$, Eq.~(14), the final
expression turns out to be strictly local:
\begin{equation}
\rm \Gamma^2_\mu (p^\prime,p) = \frac{-3\alpha_s}{8\pi} \gamma_\mu
\Gamma(2-\omega-\sigma)~,
\end{equation}
so that
\begin{equation}
\rm \Lambda_\mu (p^\prime,p) = \frac{-\alpha_s}{3\pi} \gamma_\mu
\Gamma(2-\omega-\sigma) \left|^{\alpha\to
(1/2)^+}_{\omega\to (3/2)^+}\right.~.
\end{equation}

\section{Procedure for noncovariant four-propagator and five-propagator
integrals}

In this section, we shall demonstrate our procedure for Coulomb-gauge
integrals containing precisely two noncovariant
factors of the form $\rm 1/\vec{q}^2$ and $\rm 1/(\vec{k}-\vec{q})^2$.
We shall first tackle the four-propagator
integral, then continue with the five-propagator case, consisting of
three covariant and two noncovariant propagators.

\subsection{The four-propagator integral}

Application of split dimensional regularization to the Euclidean-space
integral I,
\begin{equation}
\rm I = \int\frac{d^{2(\omega+\sigma)}{\it
q}}{(2\pi)^{2(\omega+\sigma)}} \frac{1}{q^2(q+p)^2 \vec{q}^2
(\vec{q}+\vec{p})^2}~,
\end{equation}
yields \cite{14}
\begin{eqnarray}
\rm I \!\!&=&\!\! \rm \int^1_0 dx \int^x_0 dG \int^{1-x+G}_G dy
\int^\infty_0 dAA^3
\int\frac{d^{2\omega}\vec{q}}{(2\pi)^{2\omega}}
\int \frac{d^{2\sigma}q_4}{(2\pi)^{2\sigma}} \nonumber\\
&&\nonumber\\ 
&& \!\! \rm \cdot exp\{ -A[\vec{q}^2+2y\vec{q}\cdot\vec{p} +y\vec{p}^2
+xq_4^2 +2Gq_4p_4+Gp_4^2]\}~,\\
&&\nonumber\\
&=& \!\! \rm \frac{\pi^{\omega+\sigma}}{(2\pi)^{2(\omega+\sigma)}}
\int^1_0 dx \int^x_0 dG \int_G^{1-x+G} dy
\int^\infty_0 dAA^{3-\omega-\sigma} x^{-\sigma}~. \nonumber\\
&&\nonumber\\
&& \!\! \rm \cdot exp \{ -A [(y-y^2)\vec{p}^2 + G(1-G/x)p_4^2]\}~,
\nonumber\\
&&\nonumber\\
&=&\!\! \rm  \frac{\pi^{\omega+\sigma}\Gamma
(4-\omega-\sigma)}{(2\pi)^{2(\omega+\sigma)}} \int^1_0 dx x^{-\sigma}
\int^x_0 dG
\int^{1-x+G}_G dy H^{\omega+\sigma-4}~,
\end{eqnarray}
with
\begin{equation}
\rm H (x,G,y) \equiv y (1-y) \vec{p}^2 + G(1-G/x)p_4^2~.
\end{equation}

In order to be able to integrate over y, G and x in Eq. (19), we need to
pinpoint all possible singularities in
parameter space. To this effect, we observe that the integral I {\it
diverges} for H = 0, i.e. for the following two
cases
\cite{1}:
\setcounter{equation}{0}
\def\theequation{21\alph{equation}}

\begin{eqnarray}
\rm Case~1: &\!\! \rm H \to 0,~if~y = 0~and~G = 0~;\\
\rm Case~2: &\!\! \rm H \to 0,~if~y = 1~and~G = x~.
\end{eqnarray}
Since H approaches zero {\it linearly} in Eq. (20), we may drop the
terms proportional to $\rm y^2$ and G$^2$ in Case 1,
so that

\setcounter{equation}{0}
\def\theequation{22\alph{equation}}
\begin{equation}
\rm H \to H^{(1)}_0 = y \vec{p}^2 + G p_4^2~.
\end{equation}
Similarly, setting $\rm 1-y\equiv Y$, and $\rm x-G \equiv t$ in Case 2,
we may drop the terms proportional to Y$^2$ and
t$^2$, in which case
\begin{equation}
\rm H \to H_0^{(2)} = (1-y) \vec{p}^2 + (x-G) p_4^2~.
\end{equation}
Accordingly, both $\rm H_0^{(1)}$ and $\rm H^{(2)}_0$ contribute to the
pole part of I:
\def\theequation{\arabic{equation}}
\setcounter{equation}{22}
\begin{equation}
\rm I = \frac{\pi^{\omega+\sigma} \Gamma
(2-\omega-\sigma)}{(2\pi)^{2(\omega+\sigma)}} \int^1_0 dx x^{-\sigma}
\int^x_0
dG \int^{1-x+G}_G dy (H_0^{(1)} + H_0^{(2)})^{\omega+\sigma-4} +
Finite~Terms~,
\end{equation}
or, since
%\begin{equation}
%\rm [\int dy (...)]_
%{\begin{array}{l}
%{\scriptstyle\rm y = 1}\\ \vspace*{-0.2cm}
%{\scriptstyle\rm G = x}\\
%\end{array}}
%= [\int dy (...)]_
%{\begin{array}{l}
%{\scriptstyle\rm y = 0}\\ \vspace*{-0.2cm}
%{\scriptstyle\rm G = 0}\\
%\end{array}}~,
%\end{equation}

\begin{equation}
\rm [\int dy (...)]_{{\phantom{Sum}}^{\rm y = 1}_{\rm G = x}}
= [\int dy (...)]_{{\phantom{Sum}}^{\rm y = 0}_{\rm G = 0}}~,
\end{equation}

\begin{equation}
\rm I = \frac{2\Gamma (4-\omega-\sigma)}{(4\pi)^2} \int^1_0 dx
x^{-\sigma} \int^x_0 dG \int^{1-x+G}_G dy~
[~H_0^{(1)}~]^{\omega+\sigma-4} + F.T.
\end{equation}
Hence,
\begin{eqnarray}
\rm I & \rm = \displaystyle \frac{2\Gamma (4-\omega-\sigma)}{(4\pi)^2}
\int^1_0 dx
x^{-\sigma} \int^x_0 dG G^{\omega+\sigma-3} + F.T.~,\\
& \rm = \displaystyle \frac{2\Gamma (4-\omega-\sigma)}{(4\pi)^2
p^2\vec{p}^2} \Gamma
(\omega+\sigma-2)
\int^1_0 dx x^{\omega-2} + F.T.~,
\end{eqnarray}
or, finally
\begin{equation}
\rm div \int \frac{d^4q}{(2\pi)^4}
\frac{1}{q^2(q+p)^2\vec{p}^2(\vec{q}+\vec{p})^2} = \frac{-4\Gamma
(2-\omega-\sigma)}{p^2\vec{p}^2 (4\pi)^2} \left|
{\begin{array}{l}
{\scriptstyle \sigma \to (1/2)^+}\\
\\
{\scriptstyle \omega \to (3/2)^+}\\
\end{array}}
~, \right.
\end{equation}
which is the result previously given in Table A.1 of Ref. \cite{1}.
Notice that the above noncovariant integral has
both a three-dimensional and a four-dimensional nonlocality.

\subsection{The five-propagator integral}

We now turn our attention to the massive five-propagator integral I in
Euclidean space,
\begin{equation}
\rm I = \int \frac{d^4q}{(2\pi)^4} \frac{{\it q}_i}{q^2
(q-k)^2[(p+q)^2+m^2]\vec{q}^2(\vec{q}-\vec{k})^2}~, k \equiv
p^\prime-p,~i = 1,2,3~,
\end{equation}
which is seen to possess two noncovariant and three covariant
propagators. Employing a special parametrization
\cite{14}, we may re-write I as
\begin{equation}
\rm I = \left\{\int^1_0 dx \int^x_0 dG \int^{1-x+G}_G dy \int^{x-G}_0
dv\right\} \int^\infty_0 dA A^4 \int
\frac{d^4q}{(2\pi)^4}~{\it q}_i~e^{-E}~,
\end{equation}
where:
\begin{eqnarray}
\rm E \!\! &  = & \!\!\rm Ax (q_4+r_4)^2 + A(\vec{q}+\vec{r})^2 + AZ~,
\nonumber\\
\rm r_i \!\! &  =& \!\! \rm vp_i-yk_i~,~~~~r_4 = x^{-1}
(vp_4-Gk_4)~,~~~~i = 1,2,3~,\\
\rm Z &\!\! = & \!\! \rm v [p^2+m^2+2Gp_4k_4/x-v\vec{p}^2+2y\vec{p}\cdot
\vec{k}] + y (1-y)\vec{k}^2 +
G(x-G)k_4^2/x-v^2p_4^2/x~,\nonumber
\end{eqnarray}
so that
$$
\rm I = \{...\} \int^\infty_0 d AA^4 e^{-AZ} \int d^3 \vec{q}~{\it
q}_i~e^{-A(\vec{q}+\vec{r})^2} \int d{\it q}_4~
e^{-Ax(q_4+r_4)^2}~.
$$
Defining
$$
\rm Q_i \equiv {\it q}_i +r_i~,~~~~~Q_4 \equiv {\it q}_4+r_4~,~~~~~i =
1,2,3~,
$$
and implementing split dimensional regularization, we find that
\begin{eqnarray}
\rm I \!\! & = & \!\! \rm  \{...\} \int^\infty_0 dAA^4 e^{-AZ} \int
\frac{d^{2\omega}\vec{Q}}{(2\pi)^{2\omega}}
(Q_i-r_i)~e^{-A\vec{Q}^2} \int
\frac{d^{2\sigma}Q_4}{(2\pi)^{2\sigma}}~e^{-AxQ_4^2}~,\nonumber\\
& = & \!\! \rm\{...\}
\frac{\pi^{\omega+\sigma}}{(2\pi)^{2(\omega+\sigma)}} \int^\infty_0 dA
A^{4-\omega-\sigma} (-r_i)
x^{-\sigma} e^{-AZ}~,\\
& = & \!\!\rm \frac{-\Gamma(5-\omega-\sigma)}{(4\pi)^{\omega+\sigma}}
\int^1_0 dx x^{-\sigma} \int^x_0 dG
\int^{1-x+G}_G dy
\int^{x-G}_0 dv (vp_i-yk_i) Z^{\omega+\sigma-5}~,
\end{eqnarray}
where we have made use of the following formulas:
\begin{eqnarray}
& \displaystyle\int & \!\!\!\!\!\!\rm d^{2\omega}\vec{Q} ~exp
(-A\vec{Q}^2) = \pi^\omega A^{-\omega}~, A> 0~;\nonumber\\
&&\nonumber\\
& \displaystyle\int &\!\!\!\!\!\!\rm d^{2\omega}\vec{Q}~Q_i ~exp
(-A\vec{Q}^2) = 0~, i = 1,2,3~;\nonumber\\
&&\nonumber\\
&  \displaystyle\int &\!\!\!\!\!\!\rm d^{2\sigma} Q_4 ~exp (-AxQ_4^2) =
\pi^\sigma (Ax)^{-\sigma}~;\\
&&\nonumber\\
&  \displaystyle\int^\infty_0& \!\!\!\rm dA A^{4-\omega-\sigma} ~exp
(-AZ) = \Gamma (5-\omega-\sigma)
Z^{\omega+\sigma-5}~.\nonumber
\end{eqnarray}
We re-iterate that $\vec{\rm Q}$ is defined over 2$\omega$-dimensional
complex space, and Q$_4$ over
2$\sigma$-dimensional complex space.

Our next task is to extract the pole part from the integral in Eq. (33),
by locating the singularities of the
integrand in four-dimensional parameter space. Proceeding as in Section
4.1, we see that Z in Eq. (31) vanishes for the
following sets of integration parameters:
\vskip0.3cm
\begin{itemize}
\item[Case 1:]
\begin{eqnarray}
\rm Z \!\! &\to&\!\!\rm 0,~if~v = 0,~y = 0,~G =
0,~in~which~case\nonumber\\
&&\nonumber\\
\rm Z \!\! &\to&\!\! \rm Z^{(1)}_0 = v (p^2+m^2) + y \vec{k}^2 +
Gk_4^2~.
\end{eqnarray}
\item[Case 2:]
\begin{eqnarray}
\rm Z \!\!&\to &\!\!\rm 0,~if~v = 0,~y = 1,~G =
x,~in~which~case\nonumber\\
&&\nonumber\\
\rm Z \!\! &\to &\!\! \rm Z_0^{(2)} = v (p^2+m^2+2p\cdot
k)+(1-y)\vec{k}^2 + (x-G) k_4^2~.
\end{eqnarray}
To obtain $\rm Z^{(2)}_0$ in Eq. (36), it is convenient to introduce new
parameters $\rm Y \equiv 1-y$, and $\rm t
\equiv x-G$, so that

\setcounter{equation}{0}
\def\theequation{37\alph{equation}}
\begin{equation}
\rm Z \to 0,~if~ v=0, Y=0, t=0~,
\end{equation}
and
\begin{equation}
\rm Z^{(2)}_0 = v (\bar{p}^2+m^2) + Y \vec{k}^2+tk_4^2~,~\bar{p}^2
\equiv p^2 + 2p\cdot k~,
\end{equation}
which is analogous to $\rm Z^{(1)}_0$ in Eq. (35).
\def\theequation{\arabic{equation}}
\setcounter{equation}{37}
\noindent
\end{itemize}
Accordingly, the total expression for I in Eq. (33) reads:
\begin{eqnarray}
\rm I \!\!&=& \!\!\rm 
\frac{-\Gamma(5-\omega-\sigma)}{(4\pi)^{\omega+\sigma}} \int^1_0 dx
x^{-\sigma} \int^x_0 dG
\int^{1-x+G}_G dy
\int^{x-G}_0 dv (vp_i-y k_i)\nonumber\\
&&\!\!\rm \cdot [(Z^{(1)}_0)^{\omega+\sigma-5} +
(Z_0^{(2)})^{\omega+\sigma-5}]~.
\end{eqnarray}
The divergent part of the contribution from $\rm Z^{(1)}_0$, at v = 0, y
= 0, G = 0, can be shown to vanish.

To derive the pole contribution from $\rm Z^{(2)}_0$, we replace the
parameters \{x,G,y,v\} by \{x,t,Y,v\}, whence
\begin{equation}
\rm \int^1_0 dx \int^x_0 dG\int^{1-x+G}_G dy \int^{x-G}_0 dv = -\int^1_0
dx \int^x_0 dt \int^t_{1+t-x} dY \int^t_0 dv~,
\end{equation}
and
\begin{eqnarray}
\rm I\!\! &=&\!\!\rm
\frac{\Gamma(5-\omega-\sigma)}{(4\pi)^{\omega+\sigma}} \int^1_0 dx
x^{-\sigma} \int^x_0 dt
\int^t_{1+t-x} dY \int^t_0 dv (vp_i+Yk_i-k_i) \nonumber\\
&&\!\!\rm \cdot [v(\bar{p}^2+m^2) + Y \vec{k}^2 + t
k_4^2]^{\omega+\sigma-5}~.
\end{eqnarray}

Since the integration proportional to $\rm (vp_i+Yk_i)$ leads to a
finite value, Eq. (40) reduces to the expression
\begin{eqnarray}
\rm I\!\! &=&\!\! \rm 
\frac{-k_i\Gamma(5-\omega-\sigma)}{(4\pi)^{\omega+\sigma)}}
(\bar{p}^2+m^2)^{\omega+\sigma-3}
\int^1_0 dx x^{-\sigma} \int^x_0 dt \int^t_{1+t-x} dY \nonumber\\
&& \!\!\rm \cdot \int^t_0
\frac{dv}{[v+(\bar{p}^2+m^2)^{-1}(Y\vec{k}^2+tk_4^2)]^{5-\omega-\sigma}}~,
\end{eqnarray}
or, eventually, to
\begin{eqnarray}
&\rm div&\!\!\rm \int \frac{d^4q~{\it q}_i}{(2\pi)^4 q^2 (q-k)^2 [({\it
q}+p)^2+m^2]\vec{q}^2(\vec{q}-\vec{k})^2}
\nonumber\\ 
&&\nonumber\\
&& \!\!\rm = \frac{-2k_i I^*}{k^2\vec{k}^2 [(p+k)^2 +m^2]}~, \sigma \to
(1/2)^+, \omega \to (3/2)^+~,
\end{eqnarray}
where $\rm I^* \equiv \Gamma(2-\omega-\sigma)/(4\pi)^2$. The above
integral is seen to be nonlocal in both $\vec{\rm
k}^2$ and k$^2$.

The method described here is applicable to all four- and five-propagator
cases. Particularly challenging is the {\it
basic} five-propagator integral, whose pole part consists of two
distinct terms, namely
\begin{eqnarray}
&\rm div & \!\! \rm \int \frac{d^4 q}{(2\pi)^4 q^2 (q-k)^2 [(p+k)^2+m^2]
\vec{q}^2(\vec{q}-\vec{k})^2}\nonumber\\
&&\nonumber\\
 && \!\!\rm = \frac{-2I^*}{k^2\vec{k}^2} \left(\frac{1}{p^2+m^2} +
\frac{1}{(p+k)^2+m^2}\right)~.
\end{eqnarray}

\section{BRST identity in the Coulomb gauge}

It remains to convince ourselves that $\Sigma$(p) and $\Lambda_\mu$ in
Eqs.~(10) and (16) satisfy the appropriate
one-loop BRST identity. In order to do that, however, we have to know
what the `appropriate identity' really is. The
expression usually quoted in the literature has the structure
\begin{equation}
\rm (p^\prime-p)^\mu \Lambda_\mu (p^\prime,p) = -\Sigma (p^\prime)
+\Sigma (p)~,
\end{equation}
and the question is: does this identity also hold in the Coulomb gauge,
where ghosts are known to play a significant
role? The answer to this question, surprisingly, is {\it no}.

According to Taylor, the correct BRST identity also involves ghost
contributions, $\rm G_1(p^\prime,p)$ and $\rm
G_2(p^\prime,p)$, and is given by \cite{15,16}
\begin{equation}
\rm (p^\prime-p)^\mu \Lambda_\mu (p^\prime,p) + G_1 (p^\prime,p) + G_2
(p^\prime,p) = -\Sigma(p^\prime) + \Sigma (p)~.
\end{equation}

The corresponding ghost diagrams are depicted in Figs. 4a and 4b, with
solid lines denoting quarks of mass m, curly
lines gauge bosons, and with broken lines representing ghosts; $q$ is
the internal momentum, as before, and
p,p$^\prime$ are external momenta. It remains to compute the functions
$\rm G_1, G_2$.

\begin{figure}[H]
%\centering
%\includegraphics[width=11cm]{(G.L)4.eps}
\hglue3cm
\epsfig{figure=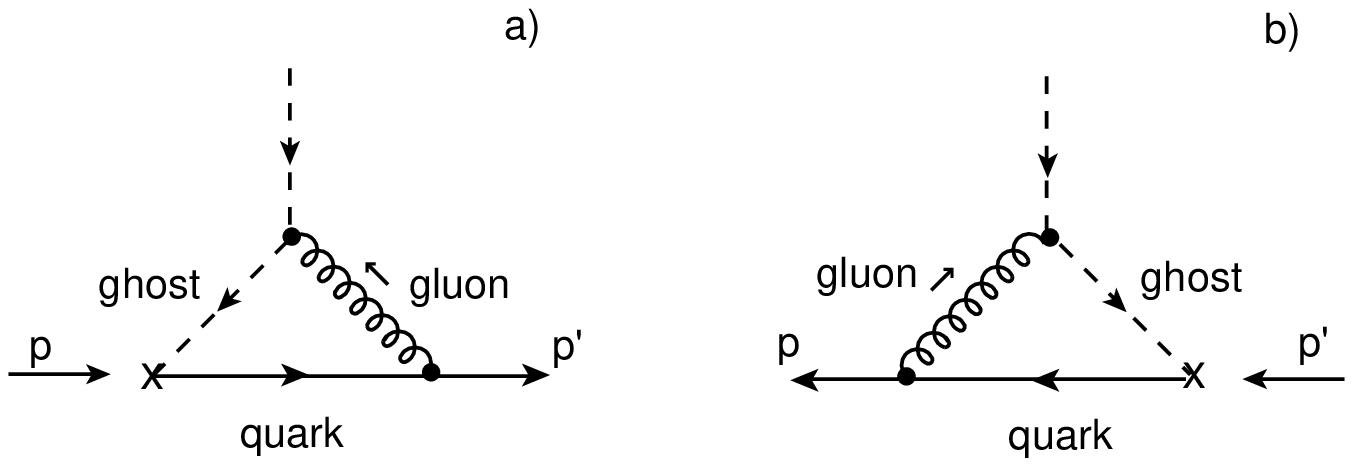,width=11cm}
\caption[]{(a) Ghost contribution \cite{15} to the IPI three-point
function, with
broken lines representing ghost particles. (b) Mirror image of the ghost
diagram
depicted in Fig. 4a.}
\end{figure}

The ghost diagram depicted in Fig. 4a leads in Minkowski space \cite{15}
to the expression $\rm G_1 (p^\prime, p)$,
\begin{eqnarray}
\rm G_1 (p^\prime, p) \!\!&=&\!\! \rm  [factors]~\int
\frac{d^4q}{(2\pi)^4} \frac{1}{[(q-k)^2+i\varepsilon]}
\left[
-\delta_{ij}+\frac{(q-k)_i(q-k)_j}{(\vec{q}-\vec{k})^2}\right]\nonumber\\
&& \!\!\rm \cdot \frac{1}{\vec{q}^2} {\it q}_i \gamma_j
\frac{1}{(\not{p}+\not{\it q} -m+i\varepsilon)} (\not{p}-m), \varepsilon
> 0, k \equiv p^\prime-p~.
\end{eqnarray}
Applying the technique of split dimensional regularization to all
Coulomb-gauge integrals, and utilizing the formulas
in the Appendix, we find that the divergent part of G$_1$ is zero, i.e.

\setcounter{equation}{0}
\def\theequation{47\alph{equation}}
\begin{equation}
\rm div G_1 (p^\prime, p) = 0~.
\end{equation}
A similar result may be established from Fig. 4b for G$_2$:
\begin{equation}
\rm div G_2 (p^\prime,p) = 0~.
\end{equation}

In summary, neither G$_1$ nor G$_2$ contributes to the BRST identity in
Eq. (45), which is obviously satisfied by
$\Sigma (p)$ and $\Lambda_\mu (p^\prime,p)$. After completion of this
work, we discovered that an identity closely
resembling the identity in Eq. (45) had been studied by Muzinich and
Paige \cite{11} and, more recently, by Newton
\cite{17} using a new approach.

\section{Genuine energy integrals at two loops}

\def\theequation{\arabic{equation}}
\setcounter{equation}{47}

The purpose of this section is to discuss briefly the application of
split dimensional regularization at the two-loop level \cite{18,19}. 
In particular, we would like to know to what extent the new 
Coulomb-gauge technique is capable of regulating both the divergences
from the notorious two-loop energy integrals \cite{5,6,7,8}, as well
as the ordinary UV divergences.  The safest way of getting at least a
partial answer to this question is to embark on an explicit two-loop
calculation and then see what happens to the ambiguous energy 
integrals.  A program of this kind has recently been initiated by the 
author, but the final results won't be available for at least a year
or two.  In the meantime, we shall confine ourselves to a few general
remarks.

Our first comment deals with the measure $\rm d^4q~d^4k$ of a 
two-loop integral.  Applying rule (1) for one-dimensional integrals,
we employ exactly two complex-dimensional parameters $\omega$ and 
$\sigma$, thus replacing the measure by
\begin{equation}
\rm d^{2(\omega+\sigma)}q~d^{2(\omega+\sigma)}k = d^{2\sigma} q_0~
    d^{2\omega} \vec{q}~d^{2\sigma} k_0~d^{2\omega} \vec{k}~;
\end{equation}
the limits $\sigma \to (1/2)^+$, and $\omega \to (3/2)^+$ are 
taken after all integrations have been executed.

Next, a general two-loop Coulomb-gauge integral is expected to give
rise to simple and double poles, proportional to
$\Gamma(2-(\omega+\sigma))$ and $[\Gamma(2-(\omega+\sigma))]^2$,
respectively; notice that $\omega$ and $\sigma$ appear additively.

We now turn our attention to a typical energy integral, such as
\cite{7}
\begin{equation}
\rm I = \int \frac {dq_0}{2\pi} \int \frac{dk_0}{2\pi}
      \,\frac {k_0}{(k_0^2-\vec{k}^2+i\epsilon)}
      \,\frac {q_0}{(q_0^2-\vec{q}^2+i\epsilon)}\,,\qquad\epsilon >0,
\end{equation}
which occurs for the first time at two loops.  According to Doust and
Taylor \cite{6,7,8}, standard dimensional regularization is incapable
of handling the ambiguities in integrals like (49).  In order to 
treat (49), Doust and Taylor use a special regularization, called
$\theta$-regularization, which assigns to (49) a particular value, 
such as $1/12$.  Our approach to this problem is somewhat different:
by applying split dimensional regularization to a certain two-loop
integral, we hope to regulate both its ordinary UV divergences, as 
well as the ambiguities from its energy integrals.  However, before
discussing a specific example, we shall briefly indicate how energy
integrals such as (49) arise in practice.

Let us consider the two-loop Yang-Mills self-energy in the Coulomb
gauge, specifically the {\it sunset diagram} (it possesses 
overlapping divergences).  The amplitude for this challenging
diagram reads as follows (we employ the notation of Ref.~\cite{20}):
\begin{equation}
\rm \Pi^{af}_{\mu\nu}(p) = \frac1{3!} \int\!\!\int 
    \frac {d^4q~d^4k}{(2\pi)^8} \,V^{abcd}_{\mu\lambda\sigma\rho}
    \,G^{dg}_{\rho\alpha}(q) \,G^{ch}_{\sigma\beta}(p-k-q)
    \,G^{be}_{\lambda\xi}(k) \,V^{ghef}_{\alpha\beta\xi\nu}~,
\end{equation}
where $\rm G^{ab}_{\mu\nu}(q)$ represents the gluon propagator in the 
Coulomb gauge, Eq. (4), and
\begin{eqnarray}
\rm V^{abcd}_{\mu\lambda\sigma\rho} \!\!&=&\!\! \rm -ig^2 
    [f^{abe} f^{cde} (\delta_{\mu\sigma} \delta_{\lambda\rho} -
                      \delta_{\mu\rho} \delta_{\lambda\sigma} ) 
   \nonumber \\ && \quad \rm
   +\, f^{ace} f^{bde} (\delta_{\mu\lambda} \delta_{\sigma\rho} -
                      \delta_{\mu\rho} \delta_{\lambda\sigma} )
   + f^{ade} f^{cbe}(\delta_{\mu\sigma} \delta_{\lambda\rho} -
                      \delta_{\mu\lambda} \delta_{\sigma\rho} )]
\end{eqnarray}
denotes the four-gluon vertex of zero-loop order.  Reduction of the
integrand yields, in Euclidean space,
\begin{eqnarray}
\rm \Pi^{af}_{\mu\nu}(p) \!\!&=&\! \rm (coeff.)\,\delta^{af} g^4 
     \int\!\!\int \frac {d^4q~d^4k}{q^2 k^2 (p-k-q)^2}
     M_{\mu\sigma\lambda\rho\alpha\xi\beta\nu}   \nonumber \\
   &&\!\! \times\ \rm 
   [A_{\rho\alpha\lambda\xi}(q,k;n)+B_{\rho\alpha\lambda\xi}(q,k;n)]
   \,C_{\sigma\beta}(q,k;p,n)~,
\end{eqnarray}
where:
\begin{eqnarray}
\rm  M_{\mu\sigma\lambda\rho\alpha\xi\beta\nu} \!\!&=& \rm
                 3 \delta_{\mu\sigma} \delta_{\lambda\rho} 
             \,( 2 \delta_{\alpha\xi} \delta_{\beta\nu} 
                 - \delta_{\alpha\nu} \delta_{\beta\xi}
                 - \delta_{\alpha\beta} \delta_{\xi\nu} ) \nonumber \\
 &&\!\!\!\!\! +\,3 \delta_{\mu\rho} \delta_{\lambda\sigma} 
             \,( 2 \delta_{\alpha\nu} \delta_{\beta\xi}
                 - \delta_{\alpha\xi} \delta_{\beta\nu} 
                 - \delta_{\alpha\beta} \delta_{\xi\nu} ) \nonumber \\
 &&\!\!\!\!\! +\,3 \delta_{\mu\lambda} \delta_{\sigma\rho} 
             \,( 2 \delta_{\alpha\beta} \delta_{\xi\nu} 
                 - \delta_{\alpha\xi} \delta_{\beta\nu} 
                 - \delta_{\alpha\nu} \delta_{\beta\xi} )~;
\end{eqnarray}
\begin{eqnarray}
\rm  A_{\rho\alpha\lambda\xi}(q,k;n) \!\!&=& \rm
           \delta_{\rho\alpha} \delta_{\lambda\xi} - \frac {k\cdot n\,
  \delta_{\rho\alpha} n_\lambda k_\xi} {\vec{k}^2} - \frac {k\cdot n\,
  \delta_{\rho\alpha} n_\xi k_\lambda} {\vec{k}^2} - \frac {q\cdot n\,
  \delta_{\lambda\xi} n_\rho q_\alpha} {\vec{q}^2} \nonumber \\
  &&\! \rm                                         -~\frac {q\cdot n\,
  \delta_{\lambda\xi} n_\alpha q_\rho} {\vec{q}^2} + \frac {q\cdot n\,
          k\cdot n\,(n_\rho q_\alpha + n_\alpha q_\rho)
          (n_\lambda k_\xi + n_\xi k_\lambda) } {\vec{q}^2 \vec{k}^2}~;
\end{eqnarray}
\begin{equation}
\rm C_{\sigma\beta} = \delta_{\sigma\beta} + \left( \frac {n^2
    R_\sigma R_\beta - R\cdot n\,(R_\sigma n_\beta + R_\beta n_\sigma)}
          {-\vec{R}^2} \right)~,\qquad R_\sigma \equiv (p-k-q)_\sigma~;
\end{equation}
$\rm B_{\rho\alpha\lambda\xi}$ consists of four terms, but is not
required for our purposes.  Substitution of expressions (53)-(55) into
Eq.~(52) leads to three kinds of integrals: (i) covariant integrals,
(ii) noncovariant Coulomb-gauge integrals, and finally, (iii) 
noncovariant Coulomb-gauge {\it energy} integrals.  At present, we are
only interested in extracting a few integrals in category (iii).

To locate a $k_0$ energy integral, for instance, we multiply the 
seventh term in $\rm  M_{\mu\sigma\lambda\rho\alpha\xi\beta\nu}$ by
the second term in $\rm  A_{\rho\alpha\lambda\xi}$, and then contract
the resulting expression with the $\rm n^2 R_\sigma R_\beta$-term in
$\rm C_{\sigma\beta}$.  Thus,
\begin{eqnarray}
\rm T_{\mu\nu}^{(1)} \!\!&=&\!\! \rm  ( 6 \delta_{\mu\lambda}
     \delta_{\sigma\rho} \delta_{\alpha\beta} \delta_{\xi\nu} )
     \left[ \frac {-k\cdot n\,\delta_{\rho\alpha} n_\lambda k_\xi} 
                  {\vec{k}^2} \right]
     \left( \frac {n^2 R_\sigma R_\beta} {-\vec{R}^2} \right)~,
     \nonumber \\ &=&\!\! \rm
     \frac {6\,k\cdot n\,R^2 n_\mu k_\nu} {\vec{k}^2 \vec{R}^2 }~,
\end{eqnarray}
so that the corresponding integral from Eq.~(52) reads
\begin{equation}
\rm I_{\mu\nu}^{(1)} \equiv n_\mu \int \frac {d^4q~d^4k~k\cdot n~k_\nu}
                     {q^2 k^2 \vec{k}^2 (\vec{p}-\vec{k}-\vec{q})^2}~,
\end{equation}
or, in Minkowski-space,
\begin{equation}
\rm I_{\mu\nu}^{(1)} = (\dots) n_\mu \int \frac {d^4q}{(q^2+i\epsilon)}
    \int \frac {d^3\vec{k}} { \vec{k}^2 (\vec{p}-\vec{k}-\vec{q})^2}
    \int \frac {dk_0~k_0\,k_\nu} {(k^2+i\epsilon)}~,
    \qquad k\cdot n = k_0\,, 
\end{equation}
leading to the $k_0$ energy integral
\begin{equation}
    \rm \int \frac {dk_0~k_0} {(k_0^2-\vec{k}^2+i\epsilon)}~.
\end{equation}
Similarly, multiplying the fourth term in
$\rm  M_{\mu\sigma\lambda\rho\alpha\xi\beta\nu}$ by the fourth term in
$\rm  A_{\rho\alpha\lambda\xi}$, and contracting with
$\rm n^2 R_\sigma R_\beta$ in $\rm C_{\sigma\beta}$, we find that
\begin{equation}
\rm T_{\mu\nu}^{(2)} \equiv
       6\,q\cdot n\,R^2 n_\mu q_\nu / (\vec{q}^2 \vec{R}^2 )~.
\end{equation}
The corresponding integral contains a $q_0$ energy integral of the form
\begin{equation}
    \rm \int \frac {dq_0~q_0} {(q_0^2-\vec{q}^2+i\epsilon)}~.
\end{equation}

To conclude this section, we shall illustrate how split dimensional
regularization handles energy integrals like (49),
\begin{equation}
\rm I = \int \frac {dq_0}{2\pi} \int \frac{dk_0}{2\pi}
      \,\frac {k_0}{(k_0^2-\vec{k}^2+i\epsilon)}
      \,\frac {q_0}{(q_0^2-\vec{q}^2+i\epsilon)}\,,\qquad\epsilon >0,
\end{equation}
which corresponds to the integral (1.3) in Doust \cite{7}.  Defining
first $k_0$ and $q_0$ each over $2\sigma$-dimensional space (and
labeling them $K_\mu$ and $Q_\mu$, respectively), and then defining
$\rm \vec{k}$ and $\rm \vec{q}$ each over $2\omega$-dimensional space
(labeling them $\vec{K}$ and $\vec{Q}$, respectively), we may rewrite
(62) as
\begin{equation}
{\rm I}_M = \int \frac {d^{2\sigma}Q}{(2\pi)^{2\sigma}}
            \int \frac {d^{2\sigma}K}{(2\pi)^{2\sigma}}    
      \,\frac {K_\mu}{(K^2-\vec{K}^2+i\epsilon)}
      \,\frac {Q_\mu}{(Q^2-\vec{Q}^2+i\epsilon)}\,,\qquad\epsilon >0;
\end{equation}
$K^2$ and $Q^2$ denote the squares of the $2\sigma$-dimensional
vectors $K_\mu$ and $Q_\mu$, respectively, while $\vec{K}^2$ and
$\vec{Q}^2$ are the squares of the $2\omega$-dimensional vectors
$\vec{K}$ and $\vec{Q}$, i.e.~the squares of the original 3-vectors
$\rm \vec{k}$ and $\rm \vec{q}$.  Wick-rotating ${\rm I}_M$ to Euclidean
space, we obtain
\begin{equation}
{\rm I}_E = \int \frac {d^{2\sigma}Q}{(2\pi)^{2\sigma}}
            \int \frac {d^{2\sigma}K}{(2\pi)^{2\sigma}} \,
      \frac {K_\mu}{(K^2+\vec{K}^2)}\,\frac {Q_\mu}{(Q^2+\vec{Q}^2)}~.
\end{equation}
Since each one of the integrals in (64) is now {\it well defined}, we
may apply {\it standard} dimensional regularization to get
\begin{equation}
    \int \frac {d^{2\sigma}Q}{(2\pi)^{2\sigma}} 
         \frac {Q_\mu}{(Q^2+\vec{Q}^2)}\ =0\,, \qquad
    \int \frac {d^{2\sigma}K}{(2\pi)^{2\sigma}}
         \frac {K_\mu}{(K^2+\vec{K}^2)}\ =0\,,
\end{equation}
so that $\rm I=0$.  Consequently, in {\it split} dimensional
regularization, the ambiguous energy integral (62), or (49), has the 
value zero.

\section{Conclusion}

In this paper we have carried out a second test of split dimensional
regularization by evaluating to one loop the pole
parts of the quark self-energy $\rm \Sigma (p)$, and the
quark-quark-gluon vertex $\rm \Lambda_\mu (p^\prime,p)$. The
technique of split dimensional regularization was specifically designed
to regulate Feynman integrals in the
noncovariant Coulomb gauge $\rm \vec{\bigtriangledown}\cdot\vec{A}^a (x)
=0$. Its most important single feature is the
use of two complex parameters, $\omega$ and $\sigma$, which permit us to
regulate the divergences of certain
energy-integrals.

The results in this paper may be summarized as follows.

\begin{enumerate}
\item Split dimensional regularization enables us to compute
unambiguously all relevant one-loop integrals in the
Coulomb gauge. This comment applies especially to the noncovariant four-
and five-propagator integrals appearing in the
vertex function $\rm \Lambda_\mu (p^\prime,p)$ (Section 4).

\item The conventional BRST identity in Eq. (44) should by rights be
replaced by the general BRST identity in Eq. (45).
The latter contains two new functions, G$_1$ and G$_2$, which correspond
to the ghost vertex diagrams shown in Figs. 4a
and 4b, respectively. An explicit calculation of $\rm G_1(p^\prime,p)$
and $\rm G_2(p^\prime,p)$ reveals, however, that
the pole parts of these functions are actually zero.

\item Computation of $\rm \Gamma^1_\mu (p^\prime,p)$ and $\rm
\Gamma^2_\mu (p^\prime,p)$ involves several complicated
Coulomb-gauge integrals, some of which turn out to be nonlocal.
Nevertheless, both $\Gamma_\mu^1$ and $\Gamma^2_\mu$
are strictly local and respect, together with $\Sigma$(p), the correct
BRST identity in Eq. (45).

\item Although ghosts generally play an important role in the Coulomb
gauge, they do not contribute explicitly to the
BRST identity relating $\Lambda_\mu$ and $\Sigma$(p).
\end{enumerate}

\section*{Acknowledgements}

It gives me great pleasure to thank John C. Taylor for numerous
discussions and comments concerning the Coulomb gauge,
and for his hospitality during the winter and spring of 1997 in the
Department of Applied Mathematics and Theoretical
Physics, Cambridge. I have also benefitted from conversations with Hugh
Osborn and Conrad Newton. I am most grateful to
Gabriele Veneziano and Alvaro De R\'ujula, and the secretarial staff,
for their hospitality during the summer of 1997 in
the Theoretical Physics Division at CERN, where this project was
completed. Finally, I should like to thank Jimmy
Williams for checking some of the more challenging Coulomb integrals.
This research was supported in part by the Natural
Sciences and Engineering Research Council of Canada under Grant No.
A8063.

\newpage

\section*{Appendix}

The following Coulomb-gauge Feynman integrals, defined over
2($\omega+\sigma$)-dimensional complex Euclidean space, are
useful in the evaluation of $\Sigma$(p), $\Gamma^1_\mu$ and
$\Gamma^2_\mu$ in Eqs.~(10), (13) and (15), respectively. We
employ the abbreviations
$$
\rm d q \equiv d^3\vec{q} dq_4/(2\pi)^4~,
$$
and
$$
\rm I^* \equiv \frac{\Gamma(2-\omega-\sigma)}{(4\pi)^2}\left|
{\begin{array}{l}
{\scriptstyle\sigma \to (1/2)^+}\\
\\
{\scriptstyle\omega \to (3/2)^+}\\
\end{array}}~. \right.
$$
\begin{eqnarray}
&\rm div&\!\!\rm \int\frac{dq~{\it q}_i{\it q}_j
q_4}{q^2[(q+p)^2+m^2]\vec{q}^2} = -\frac{1}{6} p_4 \delta_{ij} I^*~,
\nonumber\\ 
&&\nonumber\\
&\rm div&\!\!\rm \int\frac{dq~q_4^3}{q^2[(q+p)^2+m^2]\vec{q}^2} =
-\frac{3}{2} p_4 I^*~, \nonumber\\
&&\nonumber\\
&\rm div&\!\!\rm\int\frac{dq~{\it q}_i{\it q}_j
q_4^2}{(q+p)^2\vec{q}^2(\vec{q}+\vec{p})^2} = \delta_{ij}
\left(\frac{2}{15}\vec{p}^2 +\frac{2}{3}p_4^2\right)I^*-2p_ip_j
\left(\frac{1}{5} + p_4^2/(\vec{p}^2)\right)
I^*~,\nonumber\\
&&\nonumber\\
&\rm div&\!\!\rm \int\frac{dq}{(p^\prime+q)^2[(q+p)^2+m^2]\vec{q}^2} =
0~,\nonumber\\
&&\nonumber\\
&\rm div&\!\!\rm\int\frac{dq~{\it
q}_iq_4}{(p^\prime+q)^2[(p+q)^2+m^2]\vec{q}^2} = 0~, i =
1,2,3~,\nonumber\\
&&\nonumber\\
&\rm div&\!\!\rm\int
\frac{dq}{[(p^\prime+q)^2+m^2][(p+q)^2+m^2]\vec{q}^2} = 0~,\nonumber\\
&&\nonumber\\
&\rm div&\!\!\rm \int\frac{dq~{\it q}_i{\it
q}_j}{[(p^\prime+q)^2+m^2][(p+q)^2+m^2]\vec{q}^2} = \frac{1}{3}
\delta_{ij}
I^*~,\nonumber\\
&&\nonumber\\
&\rm div&\!\!\rm \int\frac{dq}{q^2 (p^\prime+q)^2[(p+q)^2+m^2]\vec{q}^2}
=
\frac{-2I^*}{(p^\prime)^2(p^2+m^2)}~,\nonumber\\
&&\nonumber\\
&\rm div&\!\!\rm\int\frac{dq~{\it q}_i{\it
q}_j}{q^2(p^\prime+q)^2[(p+q)^2+m^2]\vec{q}^2} = 0~,\nonumber\\
&&\nonumber\\
&\rm div&\!\!\rm\int\frac{dq~{\it q}_i{\it
q}_j}{q^2[(p^\prime+q)^2+m^2][(p+q)^2+m^2]\vec{q}^2} = 0~,\nonumber\\
&&\nonumber\\
&\rm
div&\!\!\rm\int\frac{dq}{q^2[(p+q)^2+m^2](\vec{q}+\vec{p}^\prime)^2\vec{q}^2}=\frac{-2I^*}{(\vec{p}^\prime)^2
(p^2+m^2)}~, \nonumber\\
&&\nonumber\\
&\rm div&\!\!\rm\int\frac{dq~{\it
q}_i}{q^2[(p+q)^2+m^2](\vec{q}+\vec{p}^\prime)^2\vec{q}^2} =
0~,\nonumber\\
&&\nonumber\\
&\rm div&\!\!\rm\int\frac{dq~{\it
q}_iq_4}{q^2[(p+q)^2+m^2](\vec{q}+\vec{p}^\prime)^2\vec{q}^2} =
0~,\nonumber\\
&&\nonumber\\
&\rm div&\!\!\rm\int\frac{dq~{\it
q}_iq_j}{q^2(p^\prime+q)^2[(p+q)^2+m^2](\vec{q}+\vec{p}^\prime)^2\vec{q}^2}
=
\frac{-2p^\prime_ip^\prime_j
I^*}{(p^\prime)^2(\vec{p}^\prime)^2[(p-p^\prime)^2+m^2]}~.\nonumber
\end{eqnarray}


\begin{thebibliography}{00}
\bibitem{14} The author is indebted to his student, J. Williams, for
providing him with this particularly convenient
parametrization.
\end{thebibliography}
\end{document}